\documentclass[conference]{IEEEtran}
\IEEEoverridecommandlockouts
\usepackage{times}
\usepackage{graphicx}
\usepackage{authblk}
\usepackage[cmex10]{amsmath}
\usepackage{amsthm}
\usepackage{amsmath}
\usepackage{algorithmic}
\usepackage{array}
\usepackage{bm,cite}
\usepackage{amssymb}

\usepackage[tight,footnotesize]{subfigure}
\usepackage{url,caption, color, enumerate, epsfig}

\title{Reinforcement Learning for Dynamic Resource Optimization in 5G Radio Access Network Slicing \thanks{
		This effort is supported by the U.S. Army Research Office under contract W911NF-20-P0035. The content of the information does not necessarily reflect the position or the policy of the U.S. Government, and no official endorsement should be inferred.
}}
	\author[1,2]{Yi Shi}
	\author[2]{Yalin E. Sagduyu}
	\author[1,2]{Tugba Erpek}
	\affil[1]{\normalsize Virginia Tech, Blacksburg, VA 24061, USA}
	\affil[2]{\normalsize Intelligent Automation, Inc., Rockville, MD 20855, USA}
	
\begin{document}
\newcommand{\argmax}{\arg\!\max}
\maketitle

\begin{abstract}
The paper presents a reinforcement learning solution to dynamic resource allocation for 5G radio access network slicing. Available communication resources (frequency-time blocks and transmit powers) and computational resources (processor usage) are allocated to stochastic arrivals of network slice requests. Each request arrives with priority (weight), throughput, computational resource, and latency (deadline) requirements, and if feasible, it is served with available communication and computational resources allocated over its requested duration. As each decision of resource allocation makes some of the resources temporarily unavailable for future, the myopic solution that can optimize only the current resource allocation becomes ineffective for network slicing. Therefore, a Q-learning solution is presented to maximize the network utility in terms of the total weight of granted network slicing requests over a time horizon subject to communication and computational constraints. Results show that reinforcement learning provides major improvements in the 5G network utility relative to myopic, random, and first come first served solutions. While reinforcement learning sustains scalable performance as the number of served users increases, it can also be effectively used to assign resources to network slices when 5G needs to share the spectrum with incumbent users that may dynamically occupy some of the frequency-time blocks.
\end{abstract}


\section{Introduction}
The demand for high-rate cellular communications is ever growing with emerging applications such as virtual/augmented reality and Internet of Things. \emph{5G communications} has evolved to meet the increasing user demand with its high throughput and low latency promises. For cellular communications systems prior to 5G, static allocation of the network resources has been considered to support different types of user applications. However, this solution is inefficient and falls short of meeting the desired user quality of experience (QoE) in 5G systems.

5G introduces the \emph{Radio Access Network (RAN) slicing} capability, where the physical network infrastructure is shared among mobile virtual network operators. The static allocation of resources (such as frequency, power, and computational resources) is replaced by reserving them on the fly with \emph{network slicing} based on the dynamic user demand. The use cases to be supported by network slicing are categorized as enhanced Mobile Broadband (eMBB), massive machine-type communications (mMTC), and ultra-reliable low-latency communications (urLLC) based on the throughput and latency requirements. The details on resource allocation as part of RAN slicing are not defined yet in the 3GPP standards. To address this gap, the ongoing research has focused on how to allocate the resources as part of RAN slicing \cite{KaloxylosSurvey, Foukas, Ordonez,Rost}.
\subsection{Related Work}
In RAN slicing, the complex network dynamics make the underlying network optimization problem challenging. In \cite{TommasoRAN}, both network slicing and mobile edge computing (MEC) technologies were considered and the resource allocation that was formulated to minimize the interference among different mobile virtual network operators was shown to be NP-hard. Recently, \emph{machine learning} was applied to solve the RAN optimization problem as an alternative to the model-based optimization that becomes easily intractable due to the complexity of dynamics involving resources and requests. In \cite{NakaoML}, it was shown that in-network deep learning is promising for application and device specific identification and traffic classification problems. Deep learning was also used in \cite{ThantharateML} to manage network load efficiency and network availability.

As data may not be readily available to train complex deep learning structures for resource allocation in response to network slicing requests, a model-free approach such as \emph{reinforcement learning} (RL) has emerged as a practical solution to learn from the 5G network performance and update resource allocation decisions for network slicing.
For resource allocation as part of network slicing, RL was compared to the static and round-robin scheduling methods in \cite{LiSlicingRL}. Both bandwidth and computational resources were considered in \cite{KooSlicingRL}. Resource allocation with RL was compared to heuristic, best-effort, and random methods in \cite{WangSlicingRL}. A prototype of network slicing was implemented on an end-to-end mobile network system in \cite{LiuSlicingRL}.  RL was applied in \cite{Gursoy} for power-efficient resource allocation in cloud RANs by considering multiple transmitters and receivers at the same frequency (rather than 5G time-frequency blocks).
In \cite{Ayala-Romero19}, resource allocation with RL was studied by exploiting prediction on communication requests. RL was also used extensively for resource allocation in wireless applications other than network slicing \cite{RLsurvey, V2V, QOS}.

The communication and computation resources available to be allocated, the objectives pursued for requests in terms of latency, throughput (5G rate) and priority, and the states, actions, and rewards used in RL have not been always fully or explicitly specified or addressed in the above works on 5G network slicing. Furthermore, the constraints imposed on resources due to potential spectrum utilization of incumbent signals in the same band have not been considered before.

\subsection{Contributions}

In this paper, we present the RL solution based on \emph{Q-learning} algorithm to dynamically allocate resources for 5G network slicing. The motivation is that a myopic solution that only considers resources and demands at a given instant cannot be effective in optimizing resources over a time horizon, as current resource allocation will make some resources unavailable for future, thereby coupling the current and future resource allocation problems. We consider dynamic allocation of the \emph{resource blocks (RBs)}, \emph{transmit power} and \emph{computational resources} to support \emph{downlink} communications from a 5G base station, gNodeB, to user equipments (UEs). Each network slicing request from a UE is associated with priority, throughput, CPU usage and latency (deadline) requirement, and needs to be served for a specific duration. These requests compete for frequency-time blocks, transmit powers of the gNodeB, and computational resources spent to support their applications.

In our Q-learning solution, the states correspond to the available resources that transition over time depending on how they are occupied (for granted network slicing requests) or released (for completed requests). The actions are the assignments of resources to requests. The reward is the network utility measured as the weighted sum of requests satisfied, where weights correspond to relative priorities of these requests.

We show that Q-learning successfully allocates resources over a time horizon and provides major gains in network utility compared to \emph{myopic}, \emph{random} and \emph{first come first served (FCFS)} resource allocation algorithms. As the number of UEs increases or priorities of network slicing request change over time, we show that Q-learning successfully adapts to dynamic user demands.

We further show that Q-learning can be effectively used to dynamically allocate spectrum resources in response to \emph{incumbent users coexisting with 5G users}. Since spectrum is a scarce resource, in an effort to develop better utilization of spectrum and support the increasing data rate requirements for emerging wireless applications, the Federal Communications Commission (FCC) adopted rules for shared commercial use of the $3550$-$3700$ MHz band which is also known as 3.5-GHz \emph{Citizens Broadband Radio Service (CBRS) band} \cite{FCCRule}. A three-tiered access and authorization framework is developed to accommodate the shared federal and non-federal use of the band. 5G and radar system (incumbent user) will need to coexist in these bands once 5G systems are deployed. It is required that 5G systems do not interfere with the operation of an incumbent user such as radar.

One approach to achieve this spectrum coexistence goal is to continuously sense the spectrum by the Environmental Sensing Capability (ESC) sensors and turn off the operation once an incumbent signal is detected from the commercial system perspective. However, depending on the radar bandwidth, a naive solution to pause the operation of the 5G network may be  infeasible. A better approach would be not allocating the corresponding spectrum by the Spectrum Access System (SAS) to the commercial users of 5G once an incumbent signal is detected at a specific resource block.

It is not clear how to reallocate spectrum resources to network slices in 5G in response to incumbent user dynamics that would change the availability of frequency-time blocks over time. As RL takes the time-horizon into account instead of pursuing a myopic objective, it emerges as a viable solution to reconfigure network resources in order to support spectrum coexistence applications for 5G. We show that the proposed Q-learning solution adapts to spectrum sharing dynamics successfully and  allocates resources for 5G RAN slicing effectively in response to dynamic spectrum utilization of incumbent users.

The rest of the paper is organized as follows. Section \ref{sec:sec2} describes the resources and the network slicing requests. Section \ref{sec:sec3} defines the optimization problem for dynamic resource allocation and presents the Q-learning solution. Section \ref{sec:sec4} provides the performance evaluation results. Section \ref{sec:sec5} concludes this paper.

\section{Resources and Network Slicing Requests} \label{sec:sec2}

\begin{figure}[b]
	\centering
	\includegraphics[width=\columnwidth]{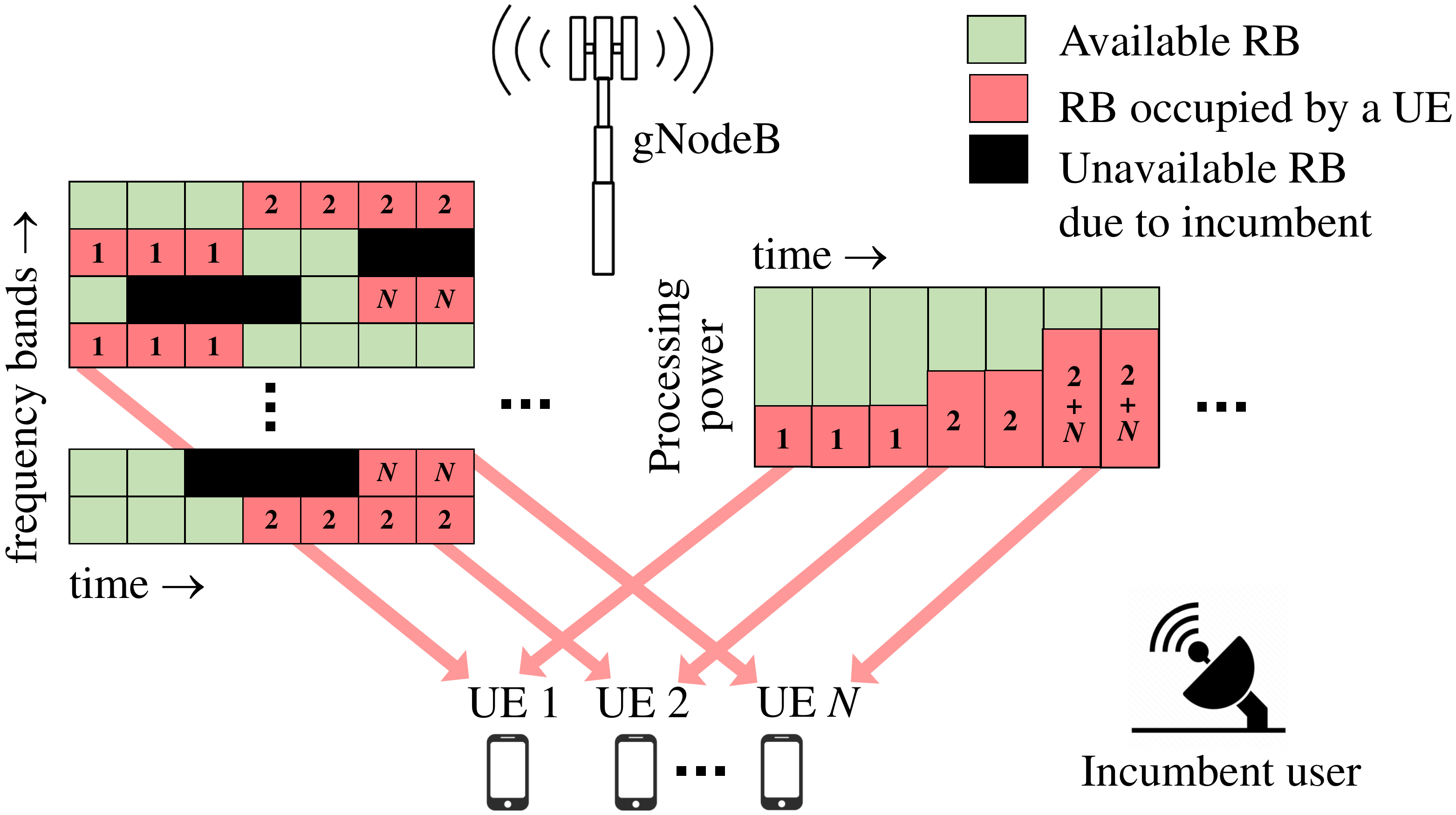}
	\caption{System model for RAN resource allocation with network slicing.}
	\label{fig:system_model}
\end{figure}

The system model is shown in Figure~\ref{fig:system_model}. Suppose that there is one gNodeB and there are $N$ UEs in a 5G network. Each UE needs to connect to the gNodeB to have network access for its applications. 5G network supports three types of traffic: eMBB, URLLC, and mMTC. Consequently, UEs may request downlink communication service (from the gNodeB to the UEs) with dynamic QoE levels regarding different throughput, CPU usage and latency (deadline) requirements, and different priorities (relative importance).

Requests are handled by the gNodeB and then appropriate network slices with corresponding resource blocks are assigned to requests. If a request is not answered yet, it stays in a queue until its deadline (time limit from the request arrival until the service starts) expires. The objective of such assignments is to maximize the weighted number of supported requests or the total provided services, where weights represent priorities of these requests. As network resources, we consider bandwidth, communication (transmit) power, and CPU usage. At time $t$, there are a set of active requests $A(t)$ that includes either requests that have just arrived or requests that are in the waiting list (i.e., requests that are not satisfied yet and their deadlines have not expired). The QoE can be measured in different forms, where we consider throughput, CPU usage, and latency
requirements. The CPU usage requirement of UE $i$ for its network slicing request $j$ is
\begin{equation} \label{eq:P_C}
P^C_{ij} \geq p^C_{ij}, \quad  (i,j) \in A(t),	
\end{equation}			
where $P^C_{ij}$ is the assigned computational resource (measured by CPU usage) and $p^C_{ij}$ is the minimum required resource.
For throughput, the QoE requirement of UE $i$ for its request $j$ is
\begin{equation} \label{eq:D}
D_{ij} \geq d_{ij}, \quad (i,j) \in A(t),	
\end{equation}	
where $D_{ij}$ is the achieved data rate and $d_{ij}$ is the minimum required rate. $D_{ij}$ is determined by the assigned bandwidth $F_{ij}$, the assigned transmit power $P^T_{ij}$ (for downlink traffic) at the gNodeB, the modulation coding scheme used for communications between the gNodeB and UE $i$, and channel effects including interference and path loss. Note that each antenna of the gNodeB serves a different user through spatial multiplexing. For 5G NR, the approximate data rate (bps) for a given number of aggregated carriers in a band or band combination is computed as follows \cite{5GNRStd1}:
\begin{equation}
r = \sum_{k=1}^K\bigg(v_{Layers}^{(k)}Q_m^{(k)}f^{(k)}R_{max}\frac{N^{BW(k),\mu}_{PRB}12}{T_s^\mu}\Big(1-O^{(k)}\Big)\bigg) , \label{eq:datarate0}
\end{equation}
where $K$ is the number of aggregated component carriers (CCs) in a band or band combination, and $R_{max} = 948/1024$. For the $k$th CC, $v_{Layers}^{(k)}$ is the maximum number of supported layers, $Q_m^{(k)}$ is the maximum supported modulation order, $f^{(k)}$ is the scaling factor that can take values $1$, $0.8$, $0.75$ or $0.4$, $\mu$ is the numerology defined in \cite{5GNRStd2}, $T_s^\mu$ is the average OFDM symbol duration in a subframe, where $T_s^\mu=\frac{10^{-3}}{14 \cdotp 2^\mu}$ for normal cyclic prefix, $N^{BW(k),\mu}_{PRB}$ is the maximum resource block allocation in the UE-supported maximum bandwidth $BW^{(k)}$ in the given band (or combination), and $O^{(k)}$ is the overhead (equal to $0.08$ for the uplink in frequency range 1). Assuming a single antenna UE with QPSK modulation, $60$ kHz subcarrier spacing and $10$ MHz bandwidth, (\ref{eq:datarate0}) becomes $r = c \cdot K$, where constant $c$ is approximately $12.59 \times 10^6$.

\begin{figure}[ht]
	\centering
	\includegraphics[width=0.85\columnwidth]{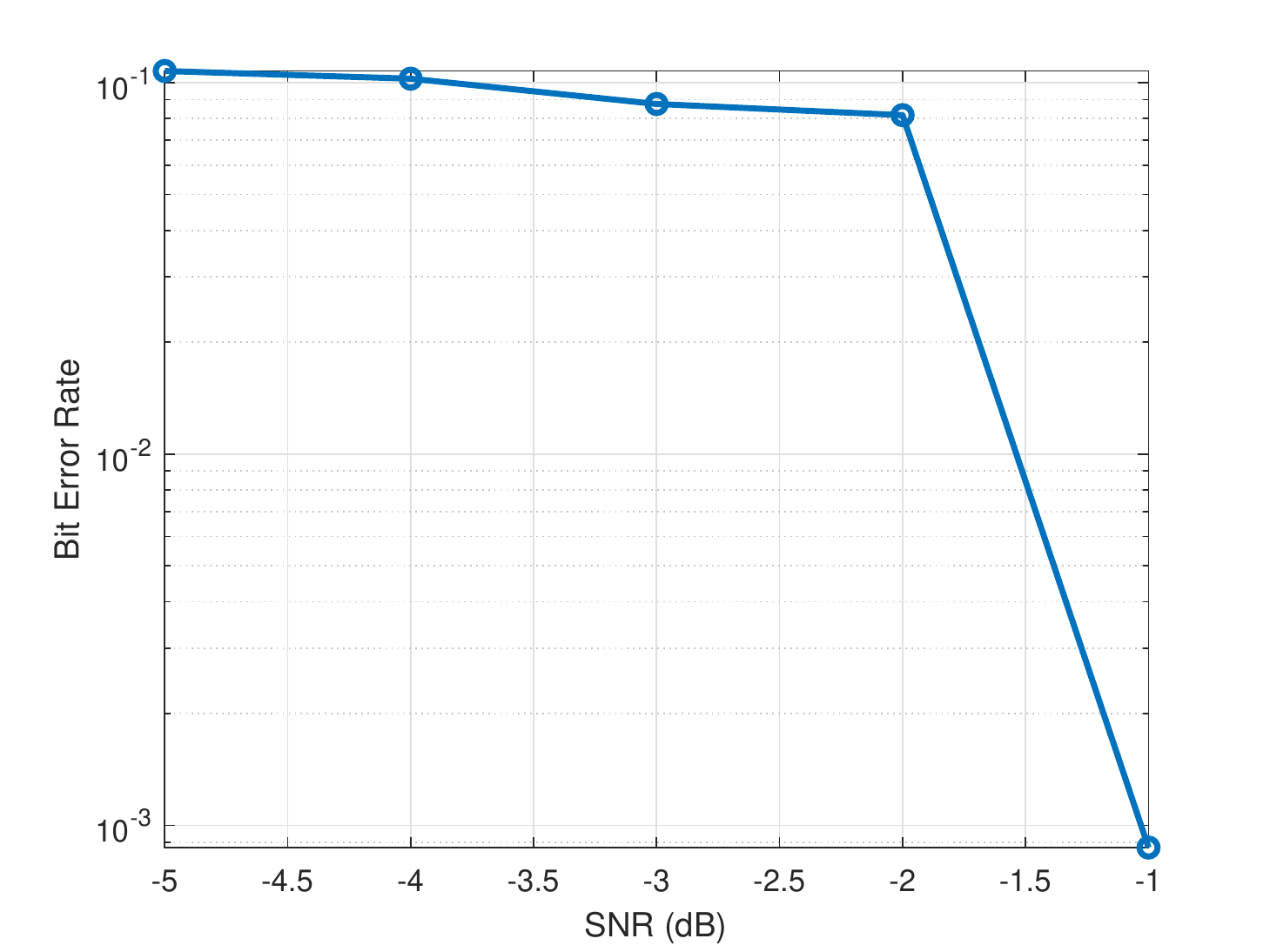}
	\caption{BER vs. SNR for QPSK signal with LDPC coding in AWGN channel.}
	\label{fig:BERvsSNR}
\end{figure}

The achieved data rate $r$ is further reduced by the corresponding bit error rate (BER) assuming that LDPC coding is used as forward error correction. To calculate the BER at different SNR levels, we simulated the performance of QPSK signal in AWGN channel with LPDC coding (see Figure~\ref{fig:BERvsSNR}). The BER value is $0$ for SNR values higher than $-1$ dB. The data rate $r$ is scaled based on the BER performance at a given SNR.
Thus, when $K_{ij}$ is the number of aggregated CCs and $\textit{BER}_{ij}$ is the BER of UE $i$ for its request $j$, (\ref{eq:D}) becomes
\begin{equation} \label{eq:D2}
c \cdot K_{ij} \cdot (1- \textit{BER}_{ij}) \geq d_{ij}, \quad (i,j) \in A(t).
\end{equation}

\section{Optimization Problem for Dynamic Resource Allocation to Network Slices} \label{sec:sec3}

Denote $x_{ij} (t)$ as the binary indicator on whether UE $i$'s request $j$ is satisfied at time $t$. The constraints of resource assignments to network slices are given as follows.
\begin{eqnarray}
\sum_{i,j} F_{ij} x_{ij}(t) & \leq & F(t), \:\: (i,j) \in A(t), \label{eq:assign-f} \\
\sum_{i,j} P^C_{ij} x_{ij}(t) & \leq & P^C(t), \:\: (i,j) \in A(t), \label{eq:assign-pc} \\
\sum_{i,j} P^T_{ij} x_{ij}(t) & \leq &  P^T(t), \:\: (i,j) \in A(t), \label{eq:assign-pt}			 
\end{eqnarray}
where $F(t)$, $P^C(t)$, and $P^T(t)$ are available communication (frequency-time blocks), computational and transmit power resources, respectively, at the gNodeB at time $t$. Some of the gNodeB's resources may have been already assigned to some requests, which are not terminated yet, and thus some resources may not be available.

The myopic objective that optimizes the resource allocation at time $t$ only is to select $F_{ij}(t)$, $P^C_{ij}(t)$ and $P^T_{ij}(t)$ for
\begin{eqnarray}
\max \sum_{ij}  w_{ij} x_{ij}(t), \:\: (i,j) \in A(t)
\label{eq:opt}
\end{eqnarray}
subject to (\ref{eq:P_C}), (\ref{eq:D2})--(\ref{eq:assign-pt}),
where $w_{ij}$ is the weight for UE $i$'s request $j$ to reflect its priority. In (\ref{eq:opt}), the myopic objective function is the reward of network slicing allocation at time $t$.

Next, we consider the optimization problem for a time horizon. The resources are updated from time $t-1$ to time $t$ as follows.
\begin{eqnarray}
F(t) & = & F(t-1)+F_r (t-1)-F_a (t-1), \label{eq:update-f} \\			
P^C(t) & = & P^C(t-1)+P^C_r (t-1)-P^C_a (t-1), \label{eq:update-pc} \\			
P^T(t) & = & P^T(t-1)+P^T_r (t-1)-P^T_a (t-1), \label{eq:update-pt}
\end{eqnarray}			
where $F_r (t-1)$, $P^C_r (t-1)$ and $P^T_r (t-1)$ are released resources at time $t-1$ on frequency, CPU usage, and transmit power, respectively, and $F_a (t)$, $P^C_a (t)$ and $P^T_a (t)$ are allocated resources at time $t$ on frequency, CPU usage and transmit power, respectively. Each request has a lifetime $l_{ij}$ and if it is satisfied at time $t$ (namely, the service starts at time $t$), this request will end at time $t+l_{ij}$. Denote $R(t)$ as the set of requests ending (completed or expired) at time $t$. The released and allocated (communications and computational) resources at time $t$ are given by
\begin{eqnarray}
F_r (t) & = & \sum_{(i,j) \in R(t)} F_{ij},	\label{eq:release-f}	\\			
P^C_r (t) & = &  \sum_{(i,j) \in R(t)} P^C_{ij}, \label{eq:release-pc} \\			
P^T_r (t) & = & \sum_{(i,j) \in R(t)} P^T_{ij}, \label{eq:release-pt}
\end{eqnarray}
and
\begin{eqnarray}
F_a (t) & = & \sum_{i,j} F_{ij}, \label{eq:allocate-f}				\\
P^C_a (t) & =&  \sum_{i,j} P^C_{ij}, \label{eq:allocate-pc}			\\
P^T_a (t) & = &  \sum_{i,j} P^T_{ij}. \label{eq:allocate-pt}
\end{eqnarray}				

Then, the optimization problem changes to
\begin{eqnarray}
\max \sum_t \sum_{ij}  w_{ij} x_{ij}(t), \:\: (i,j) \in A(t)
\label{eq:opt2}
\end{eqnarray}
subject to (\ref{eq:P_C}), (\ref{eq:D2})--(\ref{eq:assign-pt}), (\ref{eq:update-f})--(\ref{eq:allocate-pt}).
In (\ref{eq:opt2}), the objective function is the reward of network slicing allocation over a time horizon. Under an unrealistic assumption that the gNodeB knows all future requests, this problem can be solved offline. Future requests can be predicted by a deep neural network \cite{Bega19}.
Then, resources can be allocated using RL \cite{Ayala-Romero19}.
In this paper, we solve this problem by RL without any knowledge on future requests, as we will discuss in Section \ref{subsec:RLAlg}.

We also extend this formulation to dynamic spectrum sharing of 5G with incumbent legacy communication systems. One example is the CBRS band \cite{FCCRule}. The FCC authorized the use of the CBRS band for wireless service provider commercialization. In this dynamic spectrum access setting, radar is the incumbent access user (primary user) and 5G is the priority access user (secondary user). The ESC sensor needs to detect the radar signal with spectrum sensing (potentially using statistical \cite{NIST1} and machine learning \cite{NIST2} techniques). When the radar signal is detected, the SAS needs to reconfigure the 5G resource allocation to assign network slices to frequency bands that are not yet occupied by the radar. In our formulation, the available resources $A(t)$ are updated by possibly removing some frequency blocks temporarily due to radar signal occupancy.

We now consider the case that there are multiple gNodeBs, each with its own transmit power and processing power, which can be allocated independently. However, frequency resource blocks should be jointly allocated due to potential interference relationships. In addition to RL considered per gNodeB, the following pre-processing and post-processing stages need to be performed:

\begin{enumerate}
	\item \emph{Pre-processing}. A resource block is unavailable if (i) it was already allocated to some earlier requests and (ii) it was used by neighboring gNodeBs. Each gNodeB needs to sense the spectrum and identifies the resource blocks that are used by neighboring gNodeBs. Alternatively, a gNodeB can also obtain such information in the post-processing stage.
	\item \emph{Post-processing}. Neighboring gNodeBs may plan to assign the same resource blocks in their RL algorithms. To avoid interference among these assignments, a central controller can be used to resolve such conflicts (by removing some assignments) and to broadcast remaining assignments. A distributed scheme is also possible by letting each gNodeB locally broadcast its planned assignments and comparing with planned assignments from neighboring gNodeBs to decide whether a planned assignment should be kept or not.
\end{enumerate}

Once the available frequency-time blocks are determined and the unavailable ones are marked (as done in the spectrum sharing case), each gNodeB runs its RL algorithms as before to allocate its resources.

\subsection{Reinforcement Learning Algorithm} \label{subsec:RLAlg}

We use Q-learning as the model-free RL algorithm to learn the policy that determines which action (resource assignment) to take under a given state (available resources and requests) for the gNodeB. The gNodeB applies Q-learning to compute the function $Q:S\times A\to \mathbb {R}$ to evaluate the quality of action $A$ producing reward $R$ at state $S$. Note that the gNodeB maintains $Q$ as the Q-table. At each time $t$, the gNodeB selects an action $a_{t}$, observes a reward $r_{t}$, and transitions from the current state $s_t$ to a new state $s_{t+1}$ (this transition depends on current state $s_{t}$ and action $a_t$), and updates $Q$.

Starting $Q$ as a random matrix and using the weighted average of the old value and the new information, Q-learning performs the value iteration update for $Q$ as follows:
\begin{eqnarray}
\label{eq:qfunction}
Q(s_{t},a_{t}) \leftarrow && \hspace{-0.6cm} Q(s_{t},a_{t}) \\ && \hspace{-0.6cm} +  {\alpha } \cdot \left( {r_{t}} + {\gamma } \cdot \max_{a} Q(s_{t+1},a) - {Q(s_{t},a_{t})} \right), \nonumber
\end{eqnarray}
where $\alpha$ is the learning rate ($0<\alpha \leq 1$) and $\gamma$ is the discount factor ($0\leq \gamma \leq 1$) for rewards over time. In (\ref{eq:qfunction}), $\max_{a} Q(s_{t+1},a)$ refers to the estimate of the optimal future value of $Q$.

In dynamic resource allocation to network slices, the reward at time $t$ is $w_{ij}$ if UE $i$'s request $j$ is satisfied at time $t$, i.e., $x_{ij}(t) = 1$. An action of the gNodeB at time $t$ corresponds to the assignment of resources to a request at time $t$. Note that multiple actions can be taken at the same time instance.
The states at $t$ are $F(t)$, $P^C(t)$ and $P^T(t)$, namely the available resources at the gNodeB at time $t$. The transition of the state at time $t$ is driven by blocking resources for requests that are granted at time $t$ and releasing resources after the lifetimes of active services expire at time $t$. In particular, state transitions are given by (\ref{eq:update-f})-(\ref{eq:allocate-pt}).

\subsection{Baselines}
For comparison, we consider three baseline algorithms: random, first come first serve (FCFS), and myopic algorithms.
\begin{enumerate}
\item \emph{Random algorithm}: Available resources are allocated to uniformly randomly selected network slice requests.
\item \emph{FCFS algorithm}: Available resources are allocated to network slice requests based on the arrival times of requests, i.e., at any given time, the oldest network slice request is answered first provided that the available resources are sufficient to grant this request.
\item \emph{Myopic algorithm}: Available resources are allocated to maximize the current utility only by solving the optimization problem (\ref{eq:opt}).
\end{enumerate}

\section{Performance Evaluation} \label{sec:sec4}

Suppose the gNodeB receives requests possibly from three UEs. Later, we extend this setting to more UEs. For each UE, requests arrive with rate of $0.5$ per slot. Here, a slot corresponds to each time block that is $0.23$ ms long with $60$ kHz subcarrier spacing. CPU usage increases by $2\%$  increments (with total of 50 levels). For each request, weight is a random integer in [1,5], lifetime is assigned randomly in [1,10] slots, and deadline is assigned randomly in [1,20] slots. Transmit power can be at 5 levels and the maximum received SNR is selected randomly in [1.5,3]. The total frequency is $10$ MHz and is split into 11 bands. The same scenario over 1000 time slots is run to test different algorithms, namely Q-learning, random, FCFS, and myopic algorithms.
For Q-learning, we set discount factor as $\gamma=0.95$ and learning rate as $\alpha=0.1$.
The simulation code is implemented in Python.

\begin{table}[b]
	\caption{Performance comparison of Q-learning and baseline algorithms.}
	\centering
	{\small
		\begin{tabular}{c|c|c}
			Algorithm & Network utility & Ratio of improvement 
			\\ \hline \hline
			Q-learning & 1807 & -- \\ \hline
			Myopic & 1456 & 24.11\% \\ \hline
			FCFS & 1416 & 27.61\% \\ \hline
			Random & 1334 & 35.46\%
		\end{tabular}
	}
	\label{table:1}
\end{table}

Performance comparison of Q-learning and baseline algorithms is given in Table~\ref{table:1} that shows the network utility achieved by each algorithm and the ratio of network utility improvement by Q-learning relative to other algorithms.
Results show that Q-learning achieves much larger utility (total weights of satisfied requests) than other algorithms. On the other hand, myopic algorithm chooses the current best decision without consideration on future and performs better than FCFS and random resource allocations but falls far behind Q-learning. Note that all algorithms (including Q-learning and myopic algorithms) are run multiple times. The range of achieved utility is measured as $[1731,1831]$ for Q-learning and $[1359,1466]$ for myopic algorithm. These results indicate that Q-learning achieves better utility than myopic algorithm for the entire range of results.

\begin{figure}[ht]
	\centering
	\includegraphics[width=\columnwidth]{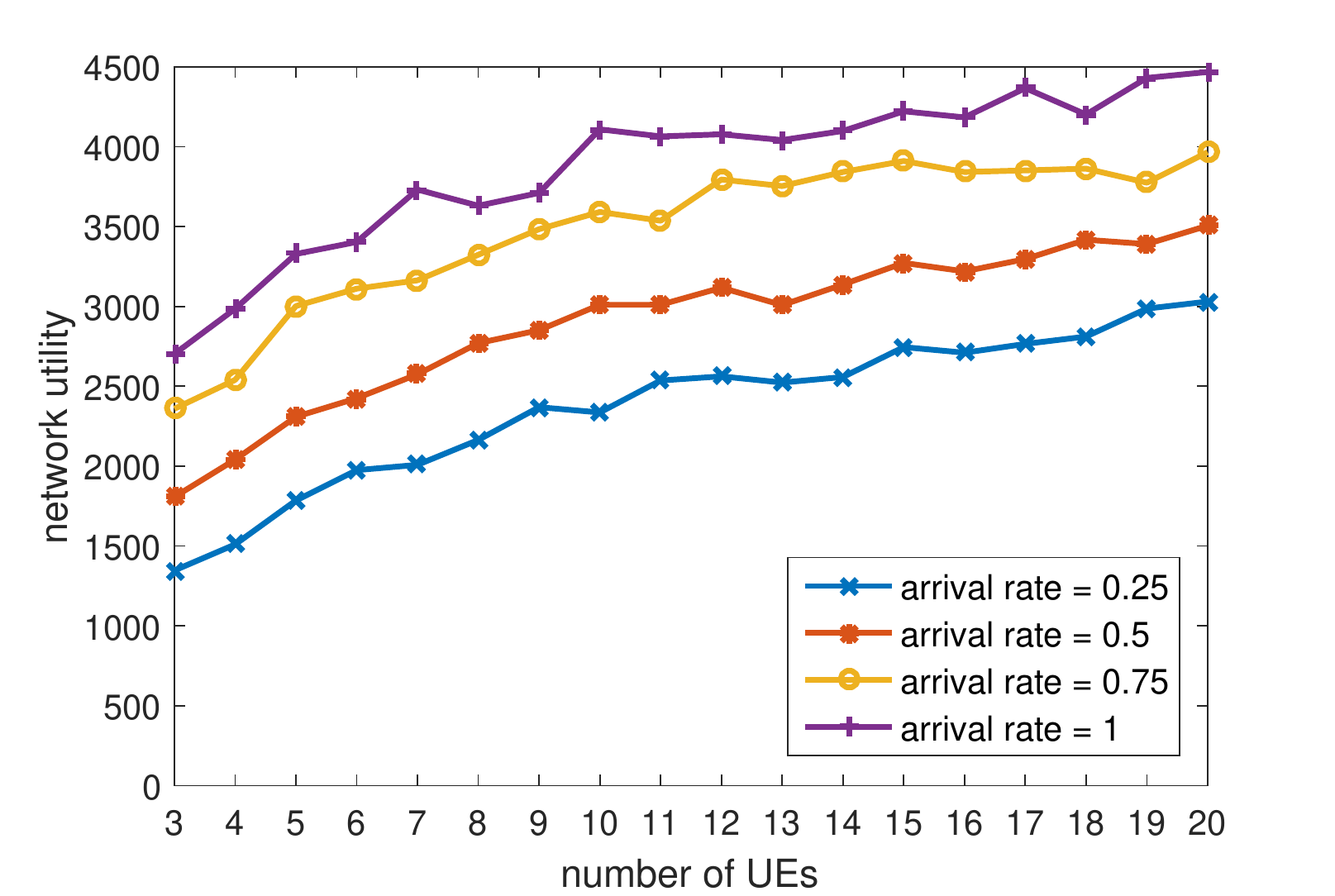}
	\caption{Network utility vs. number of UEs.}
	\label{fig:scalability}
\end{figure}

Next, we increase the number of UEs that send network slicing requests to the gNodeB and change the arrival rate at UEs. Figure~\ref{fig:scalability} shows how the network utility scales with the number of UEs when Q-learning is used.
Note that the increase of network utility with the increasing number of UEs declines as the demand for the fixed resources grows.

In optimization problem (\ref{eq:opt2}), weight $w_{ij}$ assigns priority to request $j$ of UE $i$. Next, we evaluate the effect of weights on optimization results. For that purpose, we fix the weight of all requests of one UE. The first case is [1,1,1], i.e., all requests have a weight 1. The second case is [1,3,5], i.e., all UE 1's requests have weight 1, all UE 2's requests have weight 3, and all UE 3's requests have weight 5. The third case is [5,3,1]. i.e., all UE 1's requests have weight 5, all UE 2's requests have weight 3, and all UE 3's requests have weight 1.
We show the number of served requests for each UE in Table~\ref{table:weight}.
Results indicate that if a UE's weight is increased and it is larger than others, the number of served requests for this UE increases relative other UEs. On the other hand, the actual values of weights do not play as big a role as the order of weights in terms of the effect on the network utility, i.e., weights [1,2,3] achieve the same result as weights [1,3,5].

 \begin{table}
 	\caption{The impact of weight on the number of served requests.}
 	\centering
 	{\small
 		\begin{tabular}{c|c}
 		 Weight of UEs & number of served requests \\ \hline \hline
 			[1,1,1] & [173, 139, 169] \\ \hline
 			[1,3,5] & [81, 117, 278]  \\ \hline
 			[5,3,1] & [274, 129, 68]
 		\end{tabular}
 	}
 	\label{table:weight}
 \end{table}

Next, we consider the 5G-radar spectrum coexistence scenario, where the radar signals possibly occupy multiple frequency blocks over time.
We consider two types of arrival patterns for radar signals. The first pattern corresponds to independent identically distributed (i.i.d.) arrivals and radar signal appears at any time (slot) with probability $p_I$. The second pattern corresponds to session (bursty) arrivals and the lifetime of sessions is selected uniformly randomly from [10,50] slots. Then, the arrival of sessions is adjusted to obtain $p_I$ as the probability of incumbent occupancy. We assume that the radar signal is reliably detected. In \cite{NIST2}, it was shown that the potential error in detecting radar signals becomes low when deep learning is applied.

\begin{figure}[ht]
	\centering
	\includegraphics[width=\columnwidth]{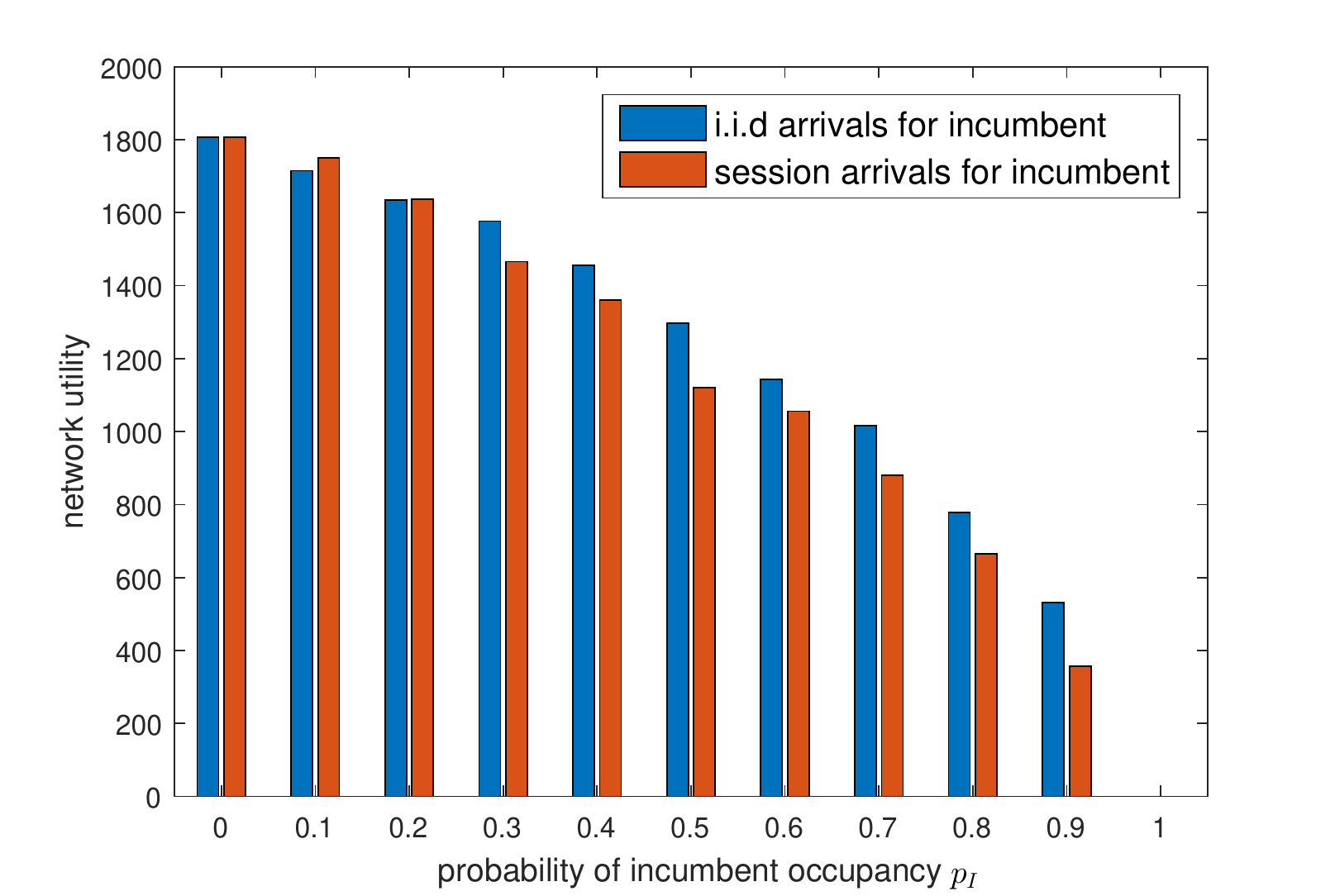}
	\caption{Network utility when 5G shares the spectrum with an incumbent user.}
	\label{fig:incumbent}
\end{figure}

In Figure~\ref{fig:incumbent}, we show the network utility achieved by Q-learning when we vary $p_I$ for both (i.i.d. and session) arrival types of incumbent user. Q-learning adapts to the spectrum occupancy pattern of the incumbent user successfully and is effective in utilizing network resources that are left from the incumbent user's spectrum occupancy. As expected, the network utility achieved by Q-learning drops as $p_I$ increases but this drop is sub-linear with $p_I$.

\section{Conclusion} \label{sec:sec5}
In this paper, we studied the problem of dynamic resource allocation for network slicing in 5G RAN and presented the RL solution that solves the underlying optimization problem over a time horizon. We considered both communication (frequency-time blocks and transmit powers) and computational resources (CPU usage) that are assigned to the incoming network slicing requests for downlink communications from the gNodeB to UEs. Each request comes with latency, rate, and CPU usage requirements. We constructed a Q-learning solution where actions are assigning resources to network slice requests and rewards are measured by the network utility, namely the weighted sum of satisfied network slicing requests (where weights account for priorities of these requests). The underlying states that transition over time correspond to the available resources that become temporarily blocked for either granted network slices or due to the spectrum use of incumbent users. As sessions of network slices are completed, their assigned resources are made available for use by other network slices. We considered random, FCFS and myopic resource allocation algorithms as baselines and showed that Q-learning outperforms them significantly in terms of the network utility. We showed that Q-learning scales well with the increasing number of UEs. We also showed that Q-learning can effectively allocate resources to network slices when some resources become unavailable randomly due to sharing the spectrum opportunistically with incumbent users.

\end{document}